\documentclass[12pt,preprint]{aastex}

\begin{document}

\Large
\centerline{\bf {Review book title:}}
\centerline{\bf { RECENT RESEARCH DEVELOPMENTS }}
\centerline{\bf {IN ASTRONOMY \& ASTROPHYSICS. }}
\vskip 3truecm
\Large
\centerline{\bf { Dark and luminous matter connections.}} 
\centerline{\bf {Towards understanding  galaxy evolution.}}
\vskip 1truecm
\centerline{Short title: { Insights into galaxy evolution}}
\vskip 3truecm
\centerline{\bf{ Paola Mazzei}}
\vskip 2truecm
\large
\centerline{INAF, Astronomical Observatory, Padova}
\centerline{ Vicolo dell' Osservatorio,5, 35122 Padova, Italy}
\centerline{e-mail: mazzei@pd.astro.it}

\normalsize

\newpage

\begin{abstract}

In this review  we outline the  major points of {\sl ab-initio} 
cosmological models focusing on some galaxy evolution unresolved problems. 
In particular our knowledge of the star formation rate which leads 
galaxy evolution, i.e.   luminosity,
colors and dust enrichment, is still very poor.
Smoothed particle hydrodynamic simulations 
including chemo-photometric evolutionary population synthesis models of isolated
collapsing systems endowed with triaxial dark matter halos, gas and the star 
formation switched on, are a useful tool to shed light on several open 
questions. This approach, which allows  a greater particle resolution than 
a fully  consistent cosmological scenario, gives us important insights into
connections between dark  and luminous matter.
Both the initial properties of the halo, as so as its geometry and dynamical 
state, and total mass of the system, drive the galaxy formation and evolution. 
In our simple framework spiral galaxies 
arise only in systems with both a total mass less than $10^{12}\,m\odot$ and a 
critical value of the  baryonic-to-total-mass ratio, around 
its cosmological value, 0.1.
For the same value of such a ratio more massive 
systems  host more luminous and  redder galaxies, as a consequence of the 
stronger burst of  star formation arising into their halos earlier on. Moreover
the star formation does not switch on in the
less massive DM halos so that the critical
mass to entail a luminous system is  $\simeq 10^{11}\,m\odot$.
For a given total mass, higher values of the baryonic-to-total-mass ratio 
provide galaxies with lower rotational support so that no constraints to the
 total mass of the  system  hosting an  elliptical galaxy have been derived. 
\end{abstract}

\section {INTRODUCTION}

Studies of the large-scale distribution of galaxies point toward understanding fundamental 
hypotheses as so as the identity of the dark matter (DM), the nature of the initial density 
perturbations and the mechanism of structure growth. Although none of these issues can be 
regarded as settled, there is now a growing consensus that cold dark matter (CDM) is the most 
likely candidate for the DM, that cosmic structure grew by the gravitational amplification of 
random phase initial density fluctuations of inflationary origin, and that the fundamental 
cosmological parameters have the following values: density parameter, 
$\Omega_0=$0.3, cosmological 
constant term, $\Lambda_0$=0.7 and  Hubble constant, $H_0$ which, in units 
of 100 km/s/Mpc, corresponds 
to  h=0.7. Cosmological constraints reflect only one aspect of the information encoded in the 
galaxy clustering. Another, equally interesting aspect, concerns the processes responsible for 
the formation and evolution of the galaxies.
The past two decades have witnessed the emergence of a successful class of {\sl 
ab-initio} theoretical models to explain the formation of the structures in the
Universe. The initial conditions 
consist of the cosmological parameters ($\Omega$, $\Lambda$, $H_0$) and of an initial fluctuation spectrum such 
as the CDM power spectrum. The remaining parameter, i.e. the amplitude of these initial 
fluctuations, is calibrated by the measured anisotropies of the microwave background. 
Numerical simulations have become an important tool for exploring the detailed predictions 
of these {\sl ab-initio} models. However, extending these studies to understand the formation of 
luminous components of galaxies is difficult since our knowledge of the star formation 
process and its interaction (feedback) with the  surrounding interstellar medium is still rather 
limited. Moreover, computational resources strongly constrain their baryonic resolution, thus 
most of the simulations that focused on the formation of individual galaxies were not 
performed in a  cosmological scenario [1, 2, 3, 4, 5, 6]. In this 
framework  [4, 6] (therein after we will refer to these papers as to paper I and II
respectively) were 
the first to present results from smoothed particle hydrodynamic  (SPH) simulations of isolated collapsing triaxial systems, 
initially composed of DM and gas, implemented with chemo-photometric evolutionary 
population synthesis (EPS) models providing the Spectral Energy Distribution (SED) from UV 
up to near-IR K band. \\
In this review we will outline the major challenges of  {\sl ab-initio}
models and our efforts  
to emphasize connections between dark and luminous matter. 
The plan of this paper is the following:  the 
next section  briefly reviews the major points  of these class of  
{\sl ab-initio} models.  Section 3  focuses on  galaxy formation in a hierarchical 
scenario and on its major problems. 
Section 4 presents our approach and points out  results from paper I and II concerning galaxy formation 
and evolution in a simple  framework, the monolithic scenario.  We will discuss the 
star formation  rate (SFR) and the rotational support (i.e. the ratio between rotational to 
kinetic energy) of the 
baryons (gas+stars) provided by these simulations and derive some issue 
concerning galaxy 
morphological types and their dependence on the properties of the whole system 
in particular  of their DM halo. 
Important findings, summarized in Section 5, are that the total mass of the system, the DM halo initial geometry,  given 
by different values of initial triaxiality ratio  and its dynamical state,  strongly reflect on 
galaxy formation and evolution. Moreover we outline the role  of different values of the 
baryonic to total mass ratio, $M_{bar}/M_{tot}$ showing that
its cosmological value, 
around 0.1, coupled with the less massive halos ($M_{tot}\le 10^{12}m\odot$), 
strongly favors formation of disk galaxies  as 
obtained in a fully consistent cosmological scenario 
[7]  where disks are the first galaxy outcome.  

\bigskip

\section{NUMERICAL SIMULATIONS OF HIERARCHICAL CLUSTERING}

During the last decade, significant progress has been made in understanding cosmic structure 
formation and evolution. Structure grows as objects of progressively larger mass merge and 
collapse to produce new virialized systems.  The formation of dark halos  in different 
cosmologies has been studied in detail with high resolution simulations [8, 9, 10, 11] 
to determine the large-scale mass 
distribution, the halo merging histories,  the structural parameters of the DM halos and finally 
the formation of galaxies inside these halos.   In many respects, these simulations are in  
agreement with observations, for example, with the present day abundance of massive galaxy 
clusters, the shape and amplitude of galaxy clustering patterns,  the magnitude of large scale 
coherent motions of galaxy systems,   among the others  [12]. However several 
difficulties still remain. Firstly, the universal cuspy density profiles of DM halos 
[9, 13, 14]  while agree with observations of 
galaxy clusters are in disagreement with flat density DM distributions inferred from 
observations  in the centers of galaxies [15, 16, 17].  As outlined by [12], 
halos formed in a $\Lambda$CDM scenario are too centrally concentrated to be consistent 
with observations of the dynamic of our own Galaxy. 
Moreover the actual physical 
mechanisms responsible for such an universal density profile are as of yet not well 
understood. [18] argue  that the universal profile does not depend crucially on 
hierarchical merging but this is a more generic feature of the 
gravitational collapse in an expanding universe which produces a near universal  
angular momentum  distribution among  the halo particles. However 
the same result is derived from SPH  simulations of isolated collapsing
triaxial systems ([4], paper I). [4] indeed argue  that such a 
profile is a consequence of violent relaxation which 
erases  all information related with the initial configuration, thus also any
cosmological signature.\\ 
Secondly, [13]  and  [19] demonstrated that  CDM scenario 
predicts that the local group should have fifty times as many dwarf galaxies than actually 
observed. In order to reconcile predictions with observations a spectrum would be required 
that suppresses power on galactic and subgalactic scales while keeping the large scale 
properties of the model virtually unchanged. This would in principle allow galaxy-sized halos 
to collapse later and thus become less centrally concentrated. However in such a case, they 
may hinder the formation of massive galaxies at high redshift, at odds with the mounting 
evidence that such  galaxies are fairly common at $z \ge 3$ [20].\\  
Thirdly, the specific angular momentum and the scale length of the disk galaxies  in the 
cosmological simulations are too small compared to real galaxies [21].
This may be partially a numerical problem of SPH  
simulations (see  Section 3), but it is more likely due to very efficient cooling and angular momentum 
transport from the baryons to the dark halos during merger events.\\ 
Forthly, in a hierarchical 
scenario  it is very difficult to fit properties of SCUBA sources detected, from
sub-millimetric surveys, at high redshift ($z>2$) and requiring  very 
high star formation rates ($>>100\, m\odot/yr$) [22, 23]. In particular they number density 
is under-predicted also by semi-analytical models [24]. 
These {\sl ab-initio}  models are based on numerical simulations as far as 
dissipationless physics is concerned however encode 
the dynamics of cooling gas, star formation,
feedback and galaxy mergers into few simple rules which allow the process of 
star formation within DM halos to be calculated [25, 26, 27].
The same models fail to reproduce the large number of high-z galaxies
detected in K-band surveys  with both Extremely Red Colors 
(EROs)  $R-K>5$,   $K \simeq 17-22$, [28, 29] and 
spectrophotometric properties consistent with   
old stellar populations  up to $z\simeq 1.5$  [30, 31].

\section{GALAXY FORMATION AND EVOLUTION}

Following the pioneering work of [32], one of the critical processes
during galaxy formation is the radiative cooling.  All the gas in virialized systems
cool at their center 
in a runaway process (over-cooling) which can be halted if a) a large fraction is 
transformed 
into stars, b) most of the gas settles in a rotationally supported disc, or c) some 
energy input reheat the gas [33].
As we will discuss with more details later on (Section 4.1.2),
since our understanding of the star formation (SF) process and its 
interaction with the surrounding interstellar medium (feedback) is still rather
limited, simulations including such a process are based on  phenomenological 
approaches.
[34] have analyzed the effect of different SF recipes in
N-body/SPH 
cosmological simulations of galaxy formation. Even if their simulations achieve very 
poor 
galaxy resolution (more than 32 particles), nevertheless these provide interesting results by  
comparing different runs. 
They conclude that results do not  depend strongly on different SFR 
prescriptions when feedback is neglected: the ratio of cold gas to stars in the final 
galaxy is 
primarily controlled by the range of gas density where the SF is allowed to proceed 
efficiently. However the  fraction of gas that cools is always much too high to account for 
the  K-band luminosity function  so  that feedback must be accounted for.
Therefore, once stars are present they are expected to return to the 
interstellar medium (ISM)  part of their mass and energy via supernovae (SNe) 
explosions, stellar winds and ultraviolet (UV)  flux. 
These latter contributions are relevant only for massive stars. 
All this is known as the stellar energy feedback.
There are  two ways of including phenomenological feedback models from type II 
supernova energy in cosmological simulations: thermal   and kinetic.
In the former case all the energy (from SNe, stellar winds and UV flux) 
supplies the thermal budget of neighboring gas particles [35, 36],
in the second one only a fraction of this energy is allowed kinetically to
affect the surrounding particles. This fraction is a free parameter [37, 38].
However thermal feedback can reduce the SFR appreciably only if  the reheated gas 
is prevented from cooling for a considerable fraction of the Hubble time.
Indeed most of the stars form in high density regions where the thermal energy
can be lost to radiation very quickly. 
Dumping the SNe energy as velocity perturbations does affect the SFR, because 
it can halt or reverse the local collapse of a region.
Therefore feedback inclusion  is a very critical point for such simulations.\\ 
Computational resources strongly constrain the baryonic resolution of numerical simulations,
thus most of the simulations that focused on the formation of individual galaxies were not 
performed in a fully cosmological scenario  [1, 2, 3, 4, 5, 39, 6], (see  also  Section 4).  
[7] present the first results of  simulations
including SF and feedback to focus on the origin of galaxy morphologies in a 
CDM Universe. Their claim for sufficiently high gas particle resolution to allow  for such a 
study. Their findings are that Hubble sequence reflects  different accretion histories in
the hierarchical framework. Disks arise from the smooth deposition of  cooled gas at 
the center of DM halos, spheroids from the stirring of preexisting disks during mergers.
Thus galaxy morphology is a transient phenomenon within the lifetime of the galaxy.
The first attempts to analyze with greater detail  formation and 
evolution of a single disk galaxy inside cosmological scenario have been performed
very recently [40, 41]. 
The general problem affecting disk galaxies in a hierarchical Universe is their higher 
concentration than  observed in spiral galaxies. This is a consequence of the formation
process of the disks. During the mergers needed to assemble their mass, the gas component, 
owing to dynamical friction, transfers most of its  angular momentum   to the surrounding
halos. Simulations show that while the specific angular momentum of the DM increases with
decreasing redshift, that of gaseous disk decreases.  So the spin of the
disks is about one order of magnitude lower than expected for spiral galaxies
and the disk scale length is shorter than observed. This produces 
 an offset in the I-band Tully-Fisher (TF) relation.
Its slope and scatter (lower than the observed 
$\simeq 0.4\, mag$) are fairly in good agreement
with the observations but its zero-point is in serious disagreement with the 
data. The simulated TF relation is almost two magnitudes faint at a given rotation 
velocity (see Fig. 1 from [12]).\\
Recent attempts to translate in photometric information the results of 
disk evolution in a 
cosmological scenario [41] derive 
a bulge-to-disk ratio 1:1  in the total (integrated) light.
The final system (z=0) rotates faster  and appears more concentrated
that observed for its mass  living 1 mag off the I-band TF relation.
\bigskip

\section{ISOLATED COLLAPSING TRIAXIAL SYSTEMS}

[4] were 
the first to present results from SPH simulations of isolated collapsing 
triaxial systems,  initially composed of DM and gas, implemented with 
chemo-photometric  evolutionary 
population synthesis (EPS) models providing the spectral energy distribution 
(SED) from UV up to near-IR K band. They have focused on the final properties 
of the  systems,  i.e. after 15 
Gyr in the rest-frame, for different values of  the stellar initial mass 
function (IMF) 
parameters (i.e.: $\phi(m)\propto m^{-\alpha}$ where $\alpha$ is slope 
and $ m_{low}$ its lower mass limit)  and feedback 
strength   self-consistently accounted for (see Section 4.2.2). In paper II,
[6]  have deepened the analysis 
focusing on the time behavior
of the SFR and on the rotational support achieved by their luminous systems.
The initial configuration is the same in both the papers
however in paper II    higher particle 
resolution runs have been  performed and a wider range of initial conditions of the 
collapsing systems have been analyzed (Table 1). 
They  aimed to shed light into 
the dependence of the system evolution on some parameters  completely 
unexplored  such as the total  mass, initial geometry and  dynamical state of 
the DM halo. Moreover also the 
effects of different values of the baryonic-to-total-mass ratio (0.1 their fiducial value)
have been investigated. 
They found important connections between dark and luminous matter.\\

\subsection{ Recipes}

\subsubsection{ The initial configuration}

As described in paper I, the system is built up with 
a density distribution i)
$\rho \propto r^{-1}$, ii) a spin parameter, $\lambda$ [42] 
given by  $|{\bf {J}}||E|^{0.5}/(GM^{0.5})$,   
where M is the total mass,
E the total energy, J the total angular momentum and G the gravitational 
constant, equal to 0.06 and aligned with the shorter principal axis of the DM 
halo, and iii) a triaxiality ratio of the DM halo as detached by
the Hubble flow in a CDM
scenario,   $\tau=(a^2-b^2)/(a^2-c^2)$ [43] where $a>b>c$, 
equal to 0.58. 
The effect of a different initial geometry, i.e. of a slightly oblate 
($\tau$=0.45) and of a prolate ($\tau$=0.84) halo, has been  investigated.
Different $\lambda$ values are also considered to  give insight into the role
of the dynamical state of the halo on the evolution of the baryonic matter.
All the simulated systems have the same initial virial ratio, 0.1, and the same
average  density to avoid a  different collapse time.
The total initial number
of particles ranges between 2000 to 20000 with $N_{DM}=N_{gas}$.
The system evolves up to 15 Gyr in the rest-frame; the final number of 
particles ranges from 10000 to 200000.\\
All the simulations performed in papers I and II include self--gravity of gas,
stars and DM,
radiative cooling, hydrodynamical pressure, shock heating, artificial viscosity,
SF and  feedback from evolved stars and type II SNe.
\medskip

\subsubsection{ The star formation rate}

As described in Section 3, the phenomenological approach for such a 
process translates in a power law dependence of the SFR on the local gas density.
In all the cases the basic procedure is the same: at the end of each time step,
a subset of gas particles is being identified as eligible to form stars. 
A fraction of them are then converted into stars which are subsequently subject
only to gravitational interactions.\\ 
The star forming criterion adopted in paper I and II
is based on following conditions [37]:
i) the local gas density must be larger than a threshold value,
$7\times 10^{-26}\, gr/cm^3$, which corresponds to the
Jeans unstability criterion and to cooling time shorter than dynamical time
in the region, and
ii) the gas particle must be in a convergent flux, i.e. $\Delta v<0$,
so pressure forces are unable to stop the collapse.
A gas particle satisfying these conditions loses a fraction
$\epsilon$ of  its mass which becomes a star particle;
$\epsilon$=0.4 is the fiducial value  however the effect of different 
choices have been also examined (0.2 in paper I and 0.04 in paper II) to account 
for  the impact of this  parameter too on the results.
\medskip

\subsubsection{ Feedback strength and initial mass function}

The feedback effects are both thermal and kinetic as described above (Sect. 3).
A fraction $f_v$ of SNe energy is supplied to
neighboring gas particles in the form of kinetic energy, with the rest,
$(1-f_v)$, being added as heat. 
Paper II assumes  $f_v=0.01$ as  fiducial value.
The kinetic energy implementation,
due to the number of massive  stars which are exploding as
SNe during each time step per stellar generation, $\Delta N_*(m>8\,m\odot)$, is
given by: $\Delta E_{SNe, vel} [erg/gr] =f_v\eta$ where
$\eta \simeq 5\times 10^{17}\,\Delta N_*(m>8\,m\odot)/M_{*}$ and $M_{*}$ is the mass
of a stellar generation which corresponds to a star particle. The value of
the ratio $\Delta N_*(m>8\,m\odot)/M_{*}$  depends on the IMF's parameters.
As presented in paper I,   B1  simulations
correspond to $\alpha=$2.5, $m_{low}=0.1\,m\odot$ and $m_{up}=100\,m\odot$
[44], B2  to  $\alpha=$2.35 with
the same  mass limits, B3  models to B2's parameters but with 
lower mass limit 0.01$\,m\odot$ [45].
Given this self-consistent approach to account for feedback, 
 B1 models provide a
feedback strength  $33\%$  higher than  B3, whereas B2 models
 $37\%$  higher than B1 and $58\%$  higher than B3  models.
\bigskip

\subsection {Results}

\subsubsection{The SFR and the rotational support of the baryonic system}

[6] find that, for a given value of the baryonic-to-total-mass ratio,
the total mass of the system, $M_{tot}$ ($10^{10}\, m\odot$, in our units),
controls the SFR. Fig.1 shows that
in systems with larger $M_{tot}$
the SF onset arises earlier on and the SFR achieves higher
values than in the  less massive halos.
In the same figure the effect of a different IMF is also shown.
B3 models allow higher SFR  as a
consequence of their lower feedback strength.
Numerical resolution slightly affects  the SFR,  the other
parameters being the same (Fig.2).
Simulations providing only  thermal feedback implementation
(i.e. $f_v=0$) 
provide a stronger burst of SFR which, as in the case of 
simulation in Fig.2 (panel ii) produces a residual gas 
fraction lower of about $24\%$ (see also Table 2, col. 9).
\begin{figure}
\plotone{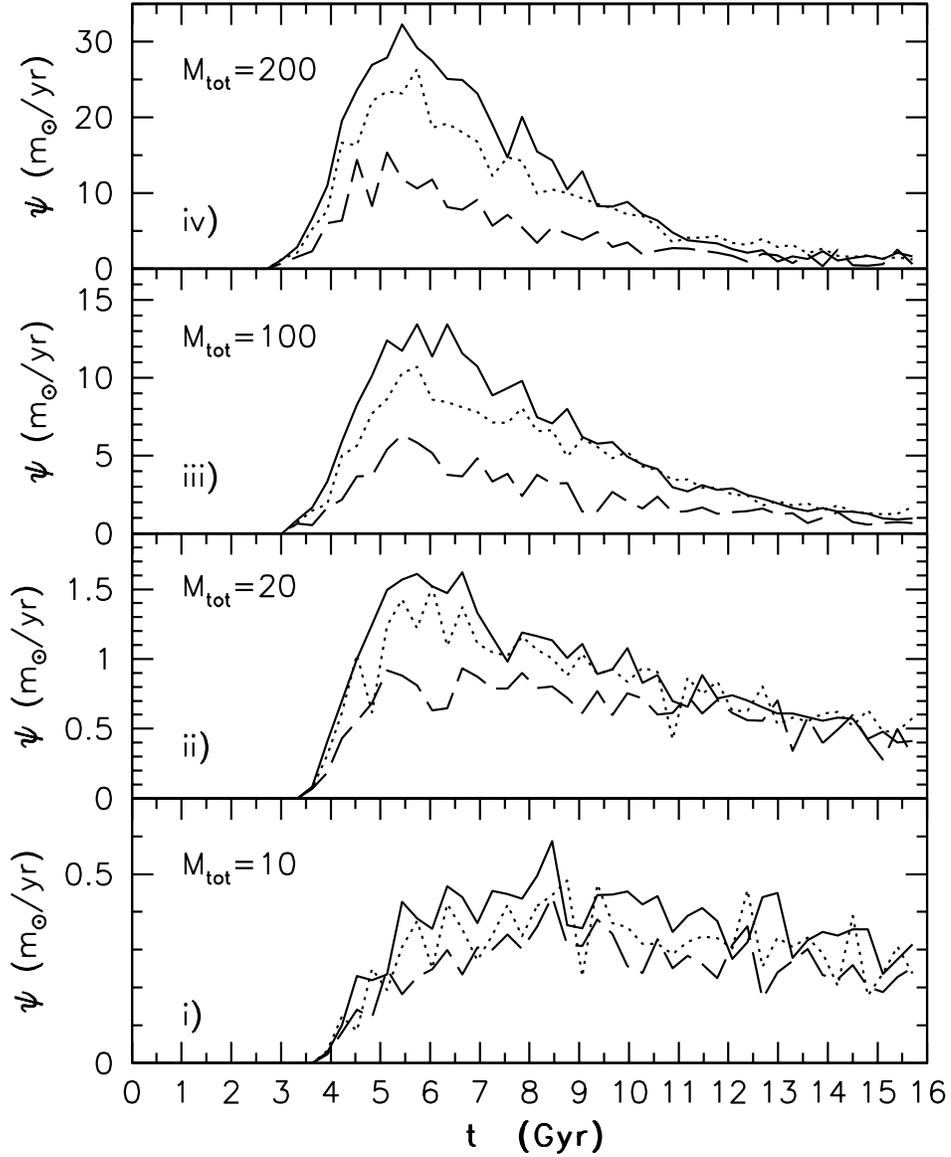}
\caption{ Time behavior of the SFR for runs 
with 6000 total initial particles,
$M_{bar}/M_{tot}$ ratio equal to 0.1, but different $M_{tot}$ and different
IMF's parameters (see text):
B1 models, dotted line, B2 models, long dashed line, B3 models,
continuous line. \label{fig1}}
\end{figure}
\begin{figure}
\plotone{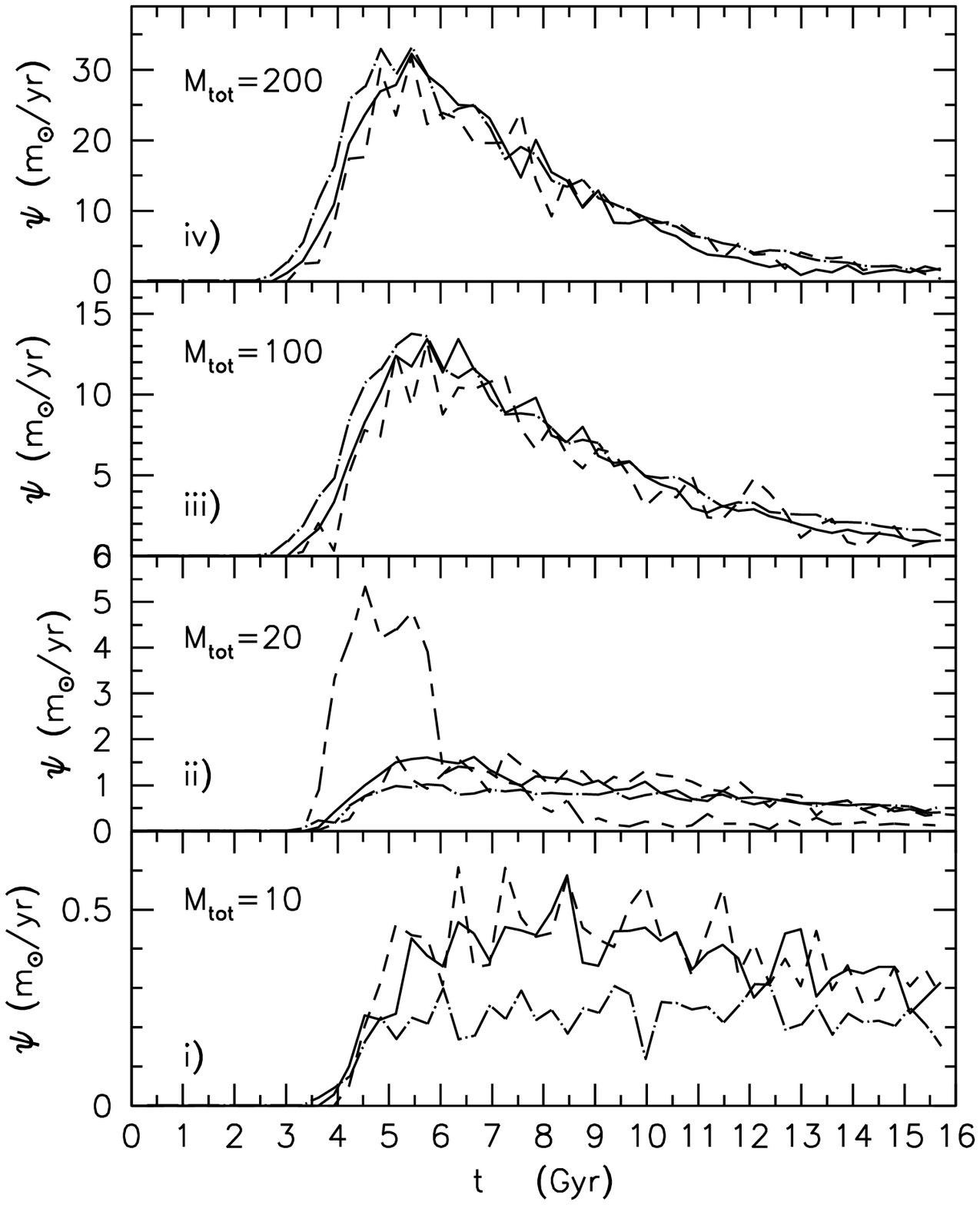}
\caption{ 
As in the previous figure  for B3 models
with different total initial particles: 6000, continuous line, 2000,
short-dashed line and 20000, dot-dashed line; in panel ii) 
a run with only thermal feedback,  i.e.$f_v=0$ (see text), and 6000 total 
initial particles is also shown  (long-short dashed line) \label{fig2}}
\end{figure}
Fig.3 (panels a, b) compares the evolution of the rotational support
of the baryons for some simulation in Fig.1 and Fig.2.
We can see that the  feedback strength, which depends on the total mass of the
system through the SFR and the IMF's parameters, strongly affects  the
dynamical evolution of the luminous matter.
This is because the lower the SFR the larger the fraction of cold ($T<10^5\,K$) rotating
gas  and the lower the amount of random motions induced by the
kinetic feedback.\\
The rotational support of the baryons increases by rising resolution
allowing the value expected for local Spirals (the maximum disk assumption
requires that such a support is $\ge 85\% \pm 10\% $ of the total energy budget [49])
for a resolution  of 6000 initial total particles.\\
\begin{figure}
\plotone{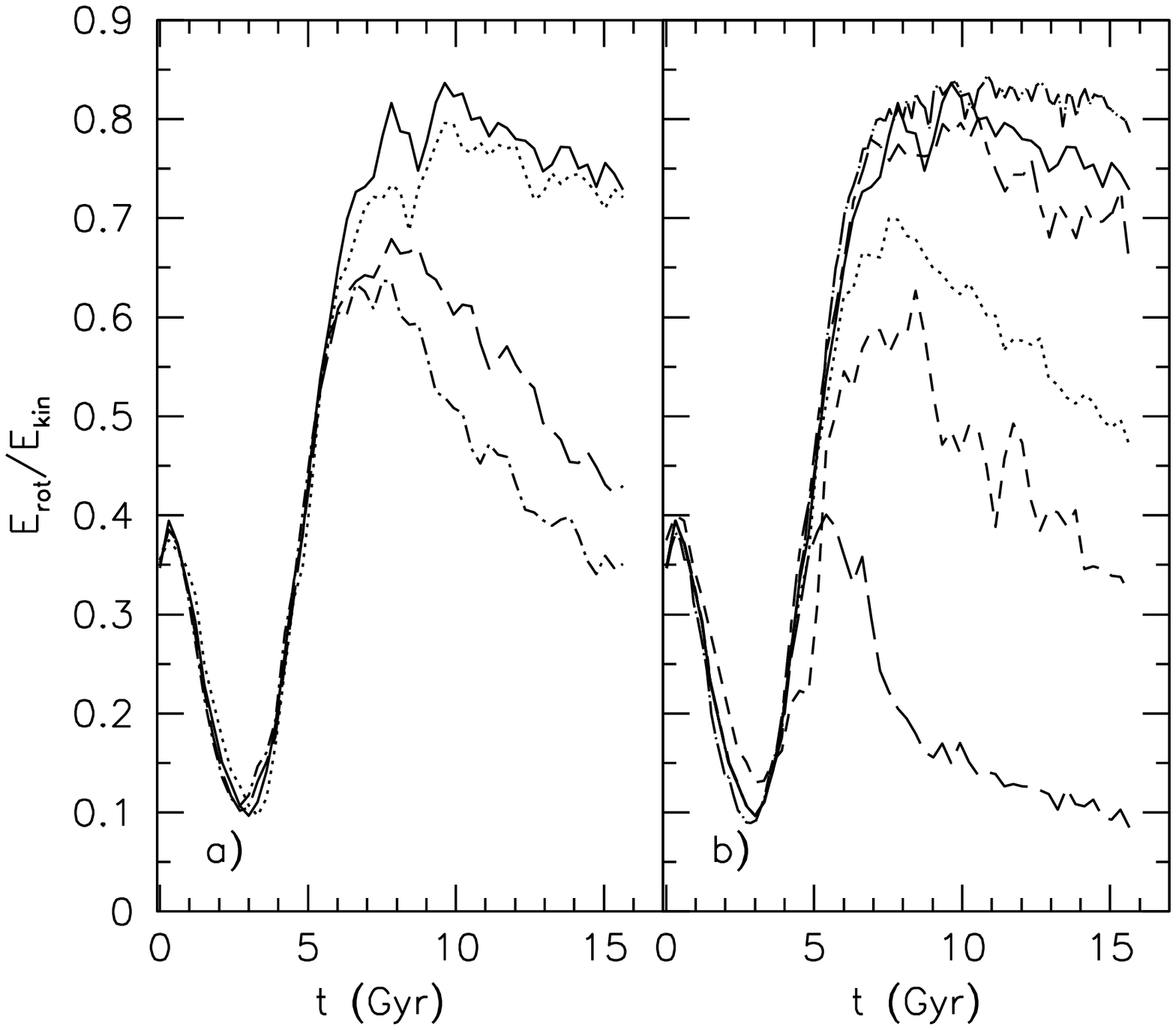}
\caption{Right panel (a) compares the evolution of the rotational support of B3 models 
in Fig.1: $M_{tot}$=10 dotted line, $M_{tot}$=20  continuous line, $M_{tot}$=100 
dashed line, $M_{tot}$=200  dot-dashed line; left panel (b) shows the same
for  models of panels ii) in  Fig.1 and  Fig.2 with the same symbols.
 \label{fig3}}
\end{figure}
Moreover even if  the SFR, and thus  the fraction of residual gas,
are  slightly depending on particle resolution, the fraction of cold gas
is decreased of about $10\%$ by reducing gas resolution from 3000 to 1000  
particles.
This is a very important point: a low resolution could strongly
affect the results of  cosmological simulations of galaxy formation provided
by such a kind of feedback.
The so called "angular momentum crisis", which
together with over-cooling problem are some  of the major challenges of
cosmological simulations, could be solved accounting for  both feedback and
high particle galaxy resolution. 
Recent findings  [40] point toward the same conclusion.
Accounting for  results of paper II (Fig.1), a suitable
choice of IMF's parameters (B3 models) improve further this point.
Moreover, as shown in Fig.3a,  
 massive baryon systems achieve lower rotational support than the
less massive ones.\\
In summary, with the fiducial value of the $M_{bar}/M_{tot}$ ratio, accounting
for the dependence of the feedback strength on the IMF's parameters claimed 
above,  B3 simulations with $M_{tot}\le 10^{12}\,m\odot$
are most suitable to match general properties of Spirals whereas more massive systems are required to match those of Ellipticals
with the  same IMF.

\subsubsection{The gas mass resolution}

The rise of the SFR with $M_{tot}$  cited above, has been further
investigated  since by keeping  constant  the  number of initial  particles
and the $M_{bar}/M_{tot}$ ratio, more massive systems entail more massive gas
clouds, so rising the mass of each gas cloud turned into stars for a given
efficiency  of SF, $\epsilon$ (see Section 4.1.2).
Thus in paper II we performed several simulations by changing $M_{tot}$ but keeping
constant $M_{bar}$ with the aim to resolve always the same mass of gas.
The gas mass resolution ranges from
$3.33\times 10^6\,m\odot$   to  $6.67\times10^7\,m\odot$; the lower resolution
is better than  that achieved both by [7], i.e. 
$1.27\times10^7\,m\odot$, and like that of [41], i.e.
$3.29\times10^6\,m\odot$. 
Results concerning B3 simulations with  $M_{bar}=1$, gas
resolution  $3.33\times 10^6\,m\odot$ and 6000 total 
initial particles, are shown in Fig. 4.
The SFR  depends on the DM mass, in the sense that it
turns on later the larger $M_{DM}$ even if the total mass of gas turned 
into stars is only slightly depending on $M_{DM}$ value.\\
\begin{figure}
\epsscale{0.5}
\plotone{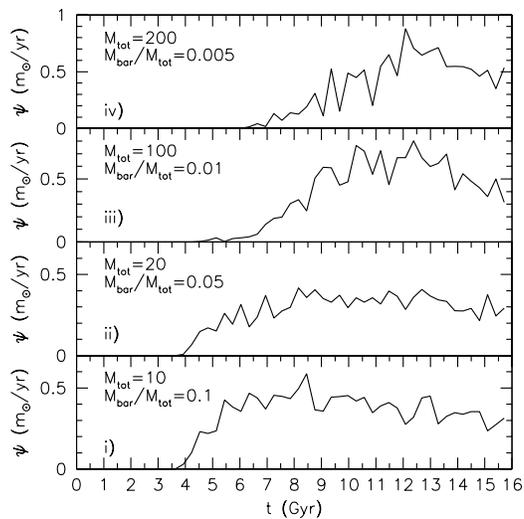}
\caption{Time behavior of the SFR for B3 models with different $M_{tot}$ but
the same gas mass resolution (see text). \label{fig4}}
\end{figure}
Looking at the dynamical evolution of the 
baryons, Fig. 5 shows that 
the  fiducial $M_{bar}/M_{tot}$ ratio (0.1) entails
the largest rotational support. The  growth of  DM mass,
which reduces this ratio well below  its fiducial value,
also reduces the rotational support of the baryons inside the system.
Thus the rotational support of the simulated galaxy is strongly
dependent also on the value of the $M_{bar}/M_{tot}$ ratio.\\
\begin{figure}
\plotone{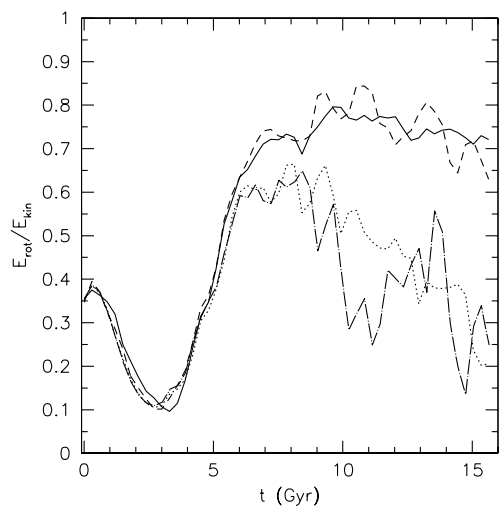}
\caption{Time behavior of the rotational support of B3 models 
in Fig.4: $M_{tot}$=10 continuous line, $M_{tot}$=20 dashed line, $M_{tot}$=100 dotted
line and $M_{tot}$=200 dotted-long dashed line. 
\label{fig5}}
\end{figure}
To conclude such a check,  some simulations have been performed with the
same $M_{bar}/M_{tot}$ ratio (0.1) but different  total masses and $\epsilon$ values
so that the same mass of each gas cloud turns on in a star particle.
Thus a simulation with $M_{tot}$ equal to 200 and
$\epsilon$ equal to 0.04 has to be compared  with  one
having  $M_{tot}=$20 and $\epsilon=$0.4 (see Fig.5 in paper II).
[6] derive that gas mass  resolution does not affect the
results strongly. For a given $M_{tot}$, the SFR does not depend on such a
resolution neither on the efficiency of SF but it is driven by the  initial
amount of gas. Systems initially gas depleted show both longer delays of the SF
onset, given the lower  gas density achieved by
the collapsing system, and lower SFRs. \\
The SF does not switch on in the
less massive baryonic systems so that, for the fiducial value of the 
$M_{bar}/M_{tot}$ ratio, the critical initial mass of gas
to entail a luminous system is  $\simeq 10^{10}\,m\odot$.

\subsubsection{ The initial geometry and dynamical state of the DM halo}

A new interesting point  emphasized by these  simulations
is that galaxy formation and
evolution is strongly affected by the initial geometry of the DM halo.
For a given $M_{tot}$ and $M_{bar}/M_{tot}$  ratio, the lower
$\tau$ the higher the SFR (Fig. 6a). This
means that less favorable conditions  to  the SFR occur inside prolate halos
($\tau>$0.66), the other parameters being the same. Moreover also
the rotational support of the galaxy depends on this  parameter (Fig.6b).
So  unfavorable conditions to spiral galaxy
formation arise inside prolate halos.
\begin{figure}
\epsscale{1}
\plotone{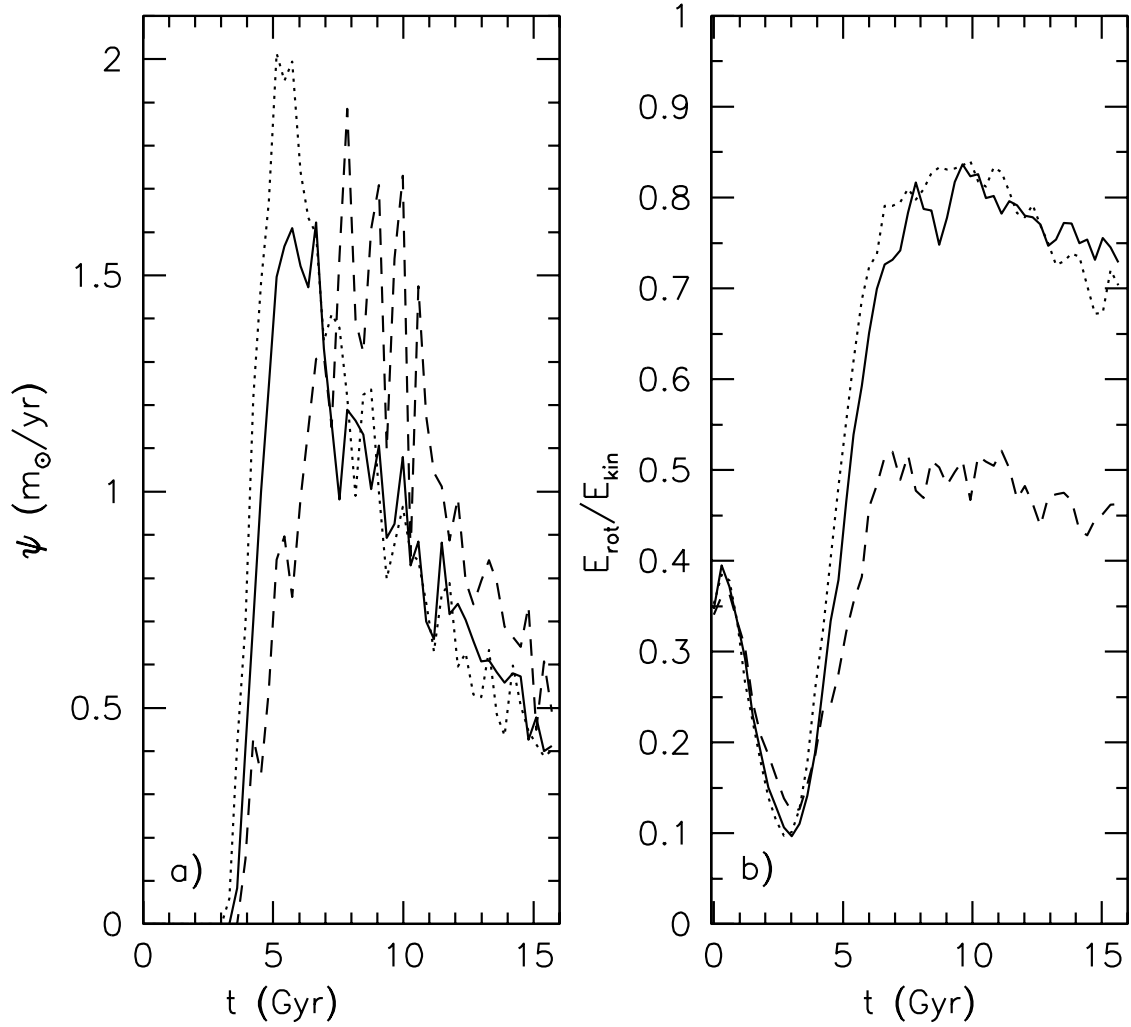}
\caption{Panel a) shows the behavior of the SFR for B3 models with
$M_{tot}$=20 and different initial triaxial ratios: continuous line
$\tau$=0.58, dotted line $\tau$=0.45 and dashed line $\tau$=0.84;
panel b)  shows  the corresponding rotational support
\label{fig6}}
\end{figure}
All the simulations  discussed so far have been built up with
initial spin parameter,  $\lambda$, equal to 0.06 (Section 4.1.1).
Simulations performed with different $\lambda$
values, to give insight into the effects of a different initial dynamical state 
on the system evolution,
show that the SF turns on later in
systems with higher  $\lambda$ and  such systems host galaxies endowed
with lower   rotational support (Fig.7a and Fig.7b respectively).
Therefore inside strongly rotating halos  spiral galaxy formation can be
suppressed.\\
\begin{figure}
\plotone{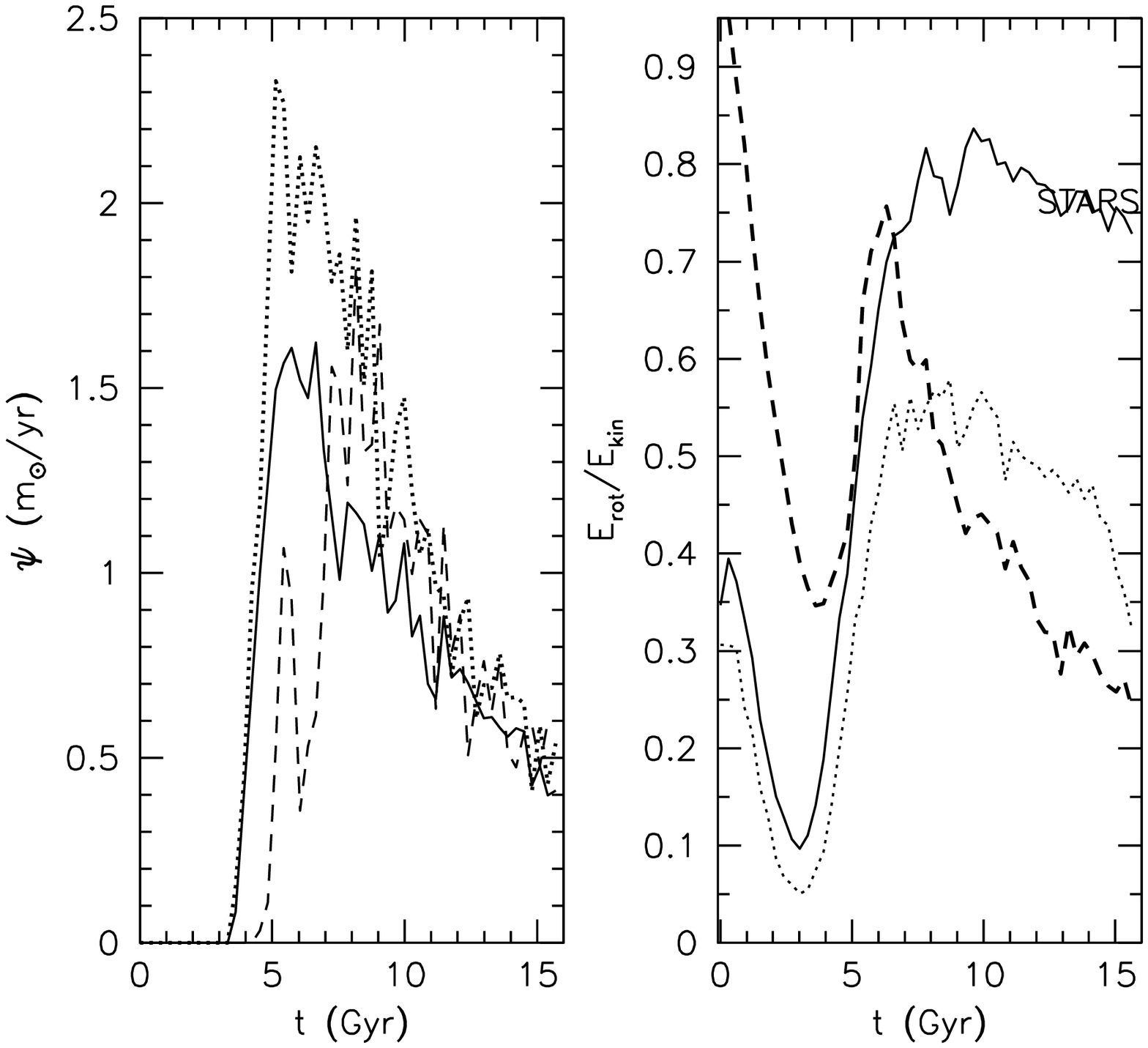}
\caption{As in Fig. 6 but for different initial spin parameters:
 continuous line
$\lambda$=0.06, dotted line $\lambda$=0.03 and dashed line $\lambda$=0.15
\label{fig7}}
\end{figure}
\subsubsection{Chemo-photometric integrated properties}
Photometric integrated galaxy properties 
depend on the SFR, for a given IMF choice [46, 47, 48].
Thus  such properties are not  affected by  different 
particle resolution  given the
slight dependence of the SFR on such a resolution, outlined above.
Integrated colors, luminosities, metallicities,
together with mass-to-light ratios, disk scale lengths and so on,
at 15 Gyr in the rest-frame, of simulations performed in paper I  are
presented in Tables 3, 4, 5, and 6 and in Fig. 17 of that paper.
Given the wider range of parameters explored in paper II (Table 1),
Fig. 8 shows  colors and metallicities of all the  simulated galaxies.
Results  are fully consistent
with observations of local galaxies of different morphological types.
\begin{figure}
\plotone{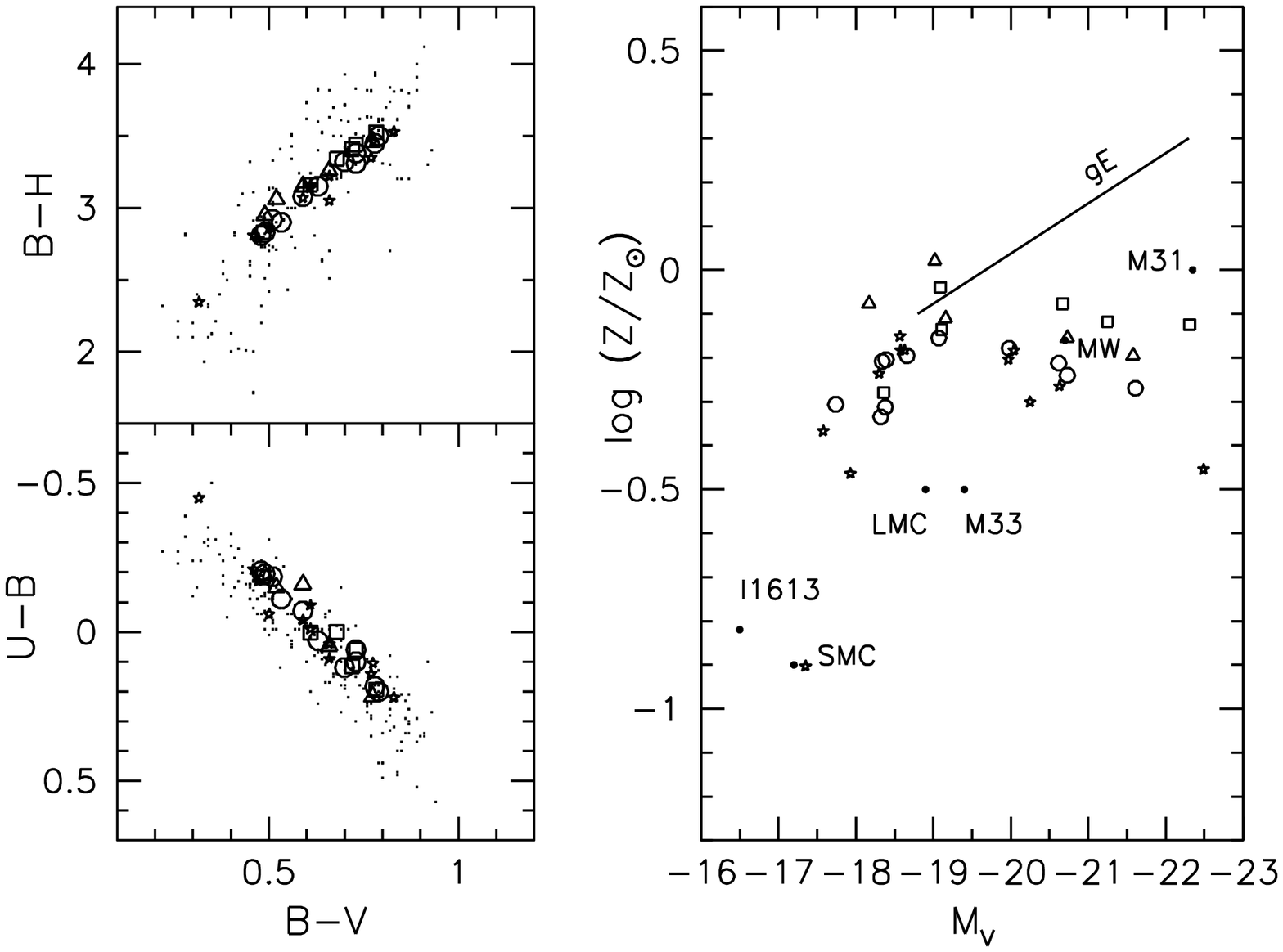}
\caption{Integrated properties at 15 Gyr in the rest-frame of simulations
in Table 1: squares are for B1, triangles for B2  and circles for
B3 models  with
$M_{tot}$=10, 20, 100, 200 and 500 (code units) and $M_{bar}/M_{tot}$=0.1;
stars are for B3 models with different initial parameters
(i.e $M_{bar}/M_{tot}$, $\tau$ and $\lambda$ values). Dots in the left panels 
are observations from [50, 51, 52], continuos line and filled circles in the 
right panel show the metallicity-luminosity relation for giant Ellipticals and
for some local systems  respectively [53].
\label{fig8}}
\end{figure}
More massive systems show redder colors, higher 
luminosity and  mass-to-light  ratios, moreover
they are characterized by older stellar populations and higher stellar 
metallicities than the less massive ones.
Furthermore the more massive
simulated galaxies have lower residual gas fraction, and in particular less 
cold gas, than the less massive ones, in
agreement with  results of fully consistent cosmological simulations 
with very low galaxy-particle resolution [34].\\
\begin{figure}
\plotone{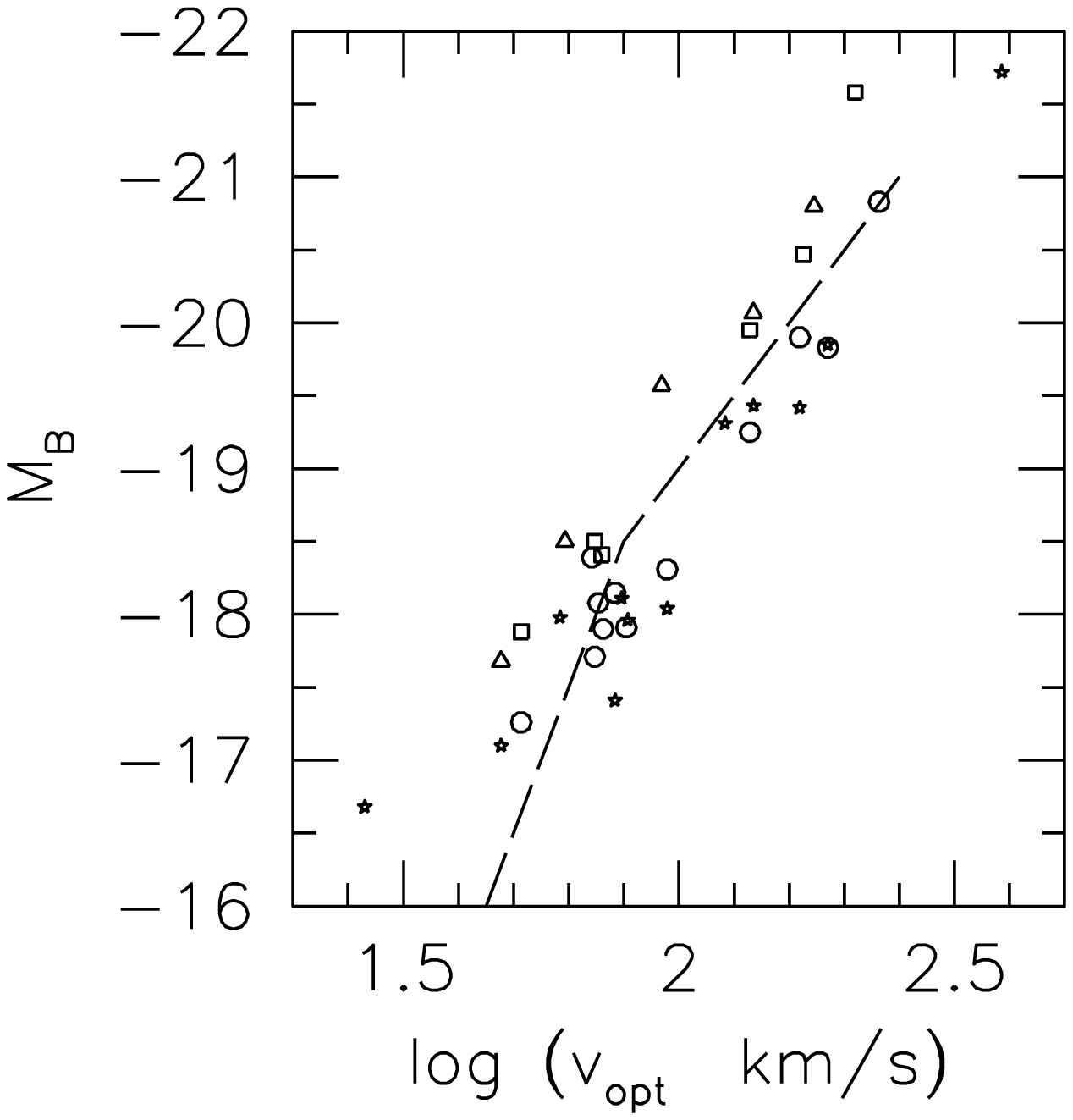}
\caption{B-band TF relation; long-dashed  line is
the best-fit of the data taken from [54] ($H_0=75\,$/km/s/Mpc); $v_{opt}$
is derived from simulations as suggested in  [54]; 
simulations are the same as in Fig. 8.
\label{fig9}}
\end{figure}
Fig.9 compares the observed B-band TF relation [54] 
with predictions of simulations in Table 1.
Table 2 summarizes disk-scale lengths and $bulge/disk$ ratios derived for
models in Fig.2b (panel ii). For an exponential disk we expect $r_{otp}=3.2r_D$
[54], in agreement with  findings in Table 2.\\
As we will discuss in the next section,
results summarized here and in the previous Sections (4.2.1-4.2.3) , 
translate in important issues concerning morphological galaxy types,
in particular  spiral galaxies, into the very
simple framework of monolithic collapse.\\

\section{CONCLUSIONS} 

Our SPH simulations of isolated collapsing triaxial systems with the star 
formation switched on [4, 6]  are aimed to shed light into the dependence of the 
system evolution on some parameters so 
far unexplored, as so as its total mass, initial dynamical state and 
geometrical shape of the 
DM halo. Further insights into the influence of different values of 
the baryonic-to-total-mass  ratio, $M_{bar}/M_{tot}$ are also given.
Important findings are that, for a given value of such a ratio, 
0.1 being the fiducial value,  the SFR depends 
on the total mass of the system and on the dynamical state of the halo whose 
initial geometry  
affects the results furthermore. These conclusions are not 
affected by gas particle resolution. Systems with larger $M_{tot}$  turn on the
SF earlier on and their SFR  achieves 
higher values than in the less massive halos. 
Moreover
the SF does not switch on in the
less massive DM halos so that the critical
mass to entail a luminous system is  $\simeq 10^{11}\,m\odot$.
According with the time behavior of the
rotational support of the baryons,   more 
suitable conditions for elliptical galaxy formation arise in the most massive 
systems while in the  less massive  ones, Spirals are  favored.
Prolate shapes of the collapsing halos, as so as higher spin parameters,  delay
the onset of the 
SF process, reduce the star formation activity and provide  baryon systems with
lower rotational support. Thus,  both the star forming activity  and the 
dynamical conditions inside these systems  favor Irr or dwarf  elliptical galaxy
formation.
Moreover  for a given value of the total mass, that of the DM
halo leads the 
morphological type of the luminous system. In particular, inside the range of
parameters explored in paper II,  only a short range of 
values of the 
baryonic-to-total-mass ratio, around 0.1, is suitable to account for Spirals: 
the higher the DM 
mass the lower the rotational support of the baryonic  system.  
Therefore we find that spiral 
galaxy formation constrains the total mass of the system: $\le 10^{12}\,m\odot$.
Accounting for the previous conclusions there are no constraints to the total mass of the 
system hosting an elliptical galaxy. However in the same framework, Ellipticals  
prefer  greater 
baryonic-to-total-mass ratios as a consequence of  both the stronger initial burst of
SF which turns  on inside them, and the lower rotational support achieved by stars.
Moreover our
findings shed light into some intriguing point of cosmological simulations
of galaxy formation. In particular if feedback effects, in term of both thermal
and kinetic energy implementation, can solve the over-cooling problem,
a careful baryon's 
resolution could allow us to overcome the so  called "angular momentum crisis" 
which is,  until now, their major challenge.\\

\section {REFERENCES}

\noindent
1. Katz, N., Hernquist, L., and Weinberg, D.H. 1992, ApJ, 399, L109.

\noindent
2. Barnes, J.,  and Hernquist, L. 1991, ApJ 370, L65.

\noindent
3. Katz, N., Weinberg, D.H., and Hernquist, L. 1996, ApJS, 105,  19.

\noindent
4. Curir, A.,  and Mazzei, P. 1999, New. Astron. 4, 1; paper I.

\noindent
5. Springel, V. 2000, MNRAS, 312, 859.

\noindent
6. Mazzei, P.,  and Curir, A. 2003, ApJ, 591, in press; paper II

\noindent
7. Steinmetz, M., and  Navarro, J.F. 2002, New Astr, 7, 155.
 
\noindent
8. Navarro, J.F.,  Frenk, C.S.,  and White, S.D.M. 1996, ApJ, 462, 563.

\noindent
9. Navarro, J.F.,   Frenk, C.S., and White, S.D.M. 1997, ApJ, 490, 493.

\noindent
10. Moore, B., Governato, F.,  Quinn, T.,Stadel, J., Lake, G. 1998, ApJ, 499, 5.

\noindent
11. Klypin, A., Kravtsov, A.V., Bullock, J.S., and Primack, J.R. 2001, ApJ, 554, 903.

\noindent
12. Steinmetz, M. 2001, IAU Symp. 208, 1, Makino, J., Hut, P., eds.

\noindent
13. Moore, B., Ghigna, S., Governato, F.,  Lake, G., et al.  1999, ApJ, 524, 
L19.

\noindent
14. Fukushige, T.,  and Makino, J. 2001, ApJ, 557, 533.

\noindent
15. Salucci, P., and Burkert, A. 2000, ApJ, 537, L9.

\noindent
16. de Blok, W.J.G., McGaugh, S.S., Bosma, A., and Rubin, V.C. 2001, ApJ, 552, L23.
 
\noindent
17. de Blok, W.J.G.,  Bosma, A., and McGaugh, S.S.  2003, MNRAS, 340, 657.

\noindent
18. Huss, A., Jain, B.,  and Steinmetz, M. 1999, ApJ, 517, 64.

\noindent
19. Klypin, A., Kravtsov, A.V., Valenzuela, O., and Prada, F. 1999, ApJ, 552, 82.

\noindent
20. Steidel, C.C., Adelberger, K.L., Dickinson, M., Giavalisco, M., et al. 1998, AJ, 
492, 428

\noindent
21. Navarro, J.F.,  and Steinmetz, M. 2000, ApJ, 538, 477.

\noindent
22. Ivinson, R., Greve, T., Smail, I., Dunlop, J.,  and Roche, N. 2002, MNRAS, 337, 1

\noindent
23. Dunne, L., Eales, S., and Edmunds, M.G. 2003, MNRAS, 341, 589.

\noindent
24. Devriendt, J.E.G., and  Guiderdoni, B. 2000, A\&A, 363, 851.

\noindent
25. White, S.D.M., and Frenk, C.S. 1991, ApJ, 379, 52.

\noindent
26. Kauffmann, G., White, S.D.M., and Guiderdoni, B. 1993, MNRAS, 264, 201.

\noindent
27. Cole, S., Aragon-Salamanca, A., Frenk, C.S., Navarro, J.F.,  and Zepf, S.E. 1994,
MNRAS, 271, 781.

\noindent
28. Cimatti, A., Pozzetti, L., Mignoli, M., et al. 2002, A\&A, 391, L1.

\noindent
29. Kashikawa, N., Takata, T., Ohyama, Y., et al. 2003, AJ, 125, 53.

\noindent
30. Daddi, E., Broadhurst, T., and Zamorani, G., et al. 2001, A\&A, 376, 825.

\noindent
31. Daddi, E., Cimatti, A., and Broadhurst, T., et al. 2002, A\&A, 384, L1.

\noindent
32. White, S.D.M., and Rees, M.J. 1978,  MNRAS, 183, 341.

\noindent
33. Navarro, J.F., and White, S.D.M. 1994, MNRAS, 267, 401.

\noindent
34. Kay, S.T., Pearce, F.R., Frenck, C.S.,  and Jenkins, A. 2002, MNRAS, 330, 113. 

\noindent
35. Katz, N. 1992, ApJ, 391, 502.

\noindent
36. Steinmetz, M.,and  Muller, E. 1994, A\&A, 281, L71.

\noindent
37. Navarro, J.F., and White, S.D.M. 1993, MNRAS, 265, 271.

\noindent
38. Mihos, J.C.,  and Hernquist, L. 1994, ApJ, 437, 611.

\noindent
39. Samland, M., and Gerhard, O.E. 2003, astro-ph/0301499
 
\noindent
40. Governato, F. Mayer, L., Wadsley, J., et al. 2003, ApjL submitted (astro-ph/0207044)
 
\noindent
41. Abadi, M.G., Navarro, J.F.,   and Steinmetz,  M. 2003, ApJ submitted (astro-ph/0211331)

\noindent
42. Barnes, J.,  and Efstathiou, G. 1987, ApJ, 319, 575.

\noindent    
43. Warren, M., S.,  Quinn, P.J., Salomon, J.K., and Zurek, W.H. 1992, ApJ, 399,  405.

\noindent
44. Miller, G.E.,  and Scalo, J.M. 1979, ApJS, 41, 513.

\noindent
45. Salpeter, E.E.  1955, ApJ, 121, 161.

\noindent
46. Sandage, A. 1986, A\&A, 161, 89.

\noindent
47. Mazzei, P., Xu, C.,  and De Zotti, G. 1992, A\&A, 256, 45.

\noindent
48. Mazzei, P.,  De Zotti, G., and Xu, C. 1994, ApJ, 426, 97.
 
\noindent
49. Sackett, P. 1997, ApJ, 483, 103.

\noindent
50. Bothun D., Aaronson, M., Schommer, B., Mould, B., Huchra, I., and Sullivan, III W.,
1985, ApJS. 57, 423.

\noindent
51.  Caldwell, N. 1983,  AJ, 88, 804.

\noindent
52. Bothun D., and Caldwell, N. 1984, ApJ, 280, 528.

\noindent
53.  Aaronson, M. 1985, in {\sl Star Forming Dwarf galaxies}, eds. D. Kanth,
T.X., Thuan and J., Tran Than Van, p.125. 

\noindent
54. Salucci, P., and Persic, M. 1997, Dark and Luminous Matter in Galaxies, APS Conf. Series, 117, 1.

\clearpage

\centerline{Table 1. Initial configuration of simulations in paper II.}
\begin{table}[hbt]
\centering{
\small
\begin{tabular}{crrrrrr}
& & &  & & &\\
N &IMF & $N_p$ & $M_{tot}$& $M_{bar}/M_{tot}$& $\tau$ & $\lambda$\\
& & &  & & &\\
1&B1 &6000 &10 &0.1 & 0.58 & 0.058\\
2&B2 &6000 &10 &0.1& 0.58& 0.058\\
3&B3 &6000 &10 &0.1& 0.58& 0.058\\
4&B3 &20000 &10 &0.1& 0.58& 0.058\\
5&B3 &6000 &10 &0.2& 0.58& 0.058\\
6&B3 &6000 &10 &0.99& 0.58& 0.058\\
7&B1 &6000 &20 &0.1 & 0.58& 0.058\\
8&B2 &6000 &20 &0.1 & 0.58& 0.058\\
9&B3 &6000 &20 &0.1 & 0.58& 0.058\\
10&B3 &20000 &20 &0.1 & 0.58& 0.058\\
11&B3 &6000 &20 &0.1 & 0.45& 0.058\\
12&B3 &6000 &20 &0.1 & 0.84& 0.058\\
13&B3 &6000 &20 &0.1 & 0.84& 0.030\\
14&B3 &6000 &20 &0.1 & 0.84& 0.150\\
15&B3 &6000 &20 &0.05 & 0.58& 0.058\\
16&B3 &6000 &20 &0.5 & 0.58& 0.058\\
17&B3 &6000 &20 &0.99 & 0.58& 0.058\\
18&B1 &6000 &100 &0.1 & 0.58& 0.058\\
19&B2 &6000 &100 &0.1& 0.58& 0.058\\
20&B3 &6000 &100 &0.1& 0.58& 0.058\\
21&B3 &20000 &100 &0.1& 0.58& 0.058\\
22&B3 &6000 &100 &0.01& 0.58& 0.058\\
23&B3 &6000 &100 &0.02& 0.58& 0.058\\
24&B3 &6000 &100 &0.99& 0.58& 0.058\\
25&B1 &6000 &200 &0.1 & 0.58& 0.058\\
26&B2 &6000 &200 &0.1& 0.58& 0.058\\
27&B3 &6000 &200 &0.1& 0.58& 0.058\\

\end{tabular}}
\end{table}
\clearpage

\centerline{Table 1: continued}
\begin{table}[hbt]
\centering{
\begin{tabular}{crrrrrr}
& & &  &  & & \\
N & IMF & $N_p$ & $M_{tot}$& $M_{bar}/M_{tot}$& $\tau$ & $\lambda$\\
& & &  & & & \\
28&B3 &20000 &200 &0.1& 0.58& 0.058\\
29&B3 &6000 &200 &0.1& 0.45& 0.058\\
30&B3 &6000 &200 &0.05& 0.45& 0.058\\
31&B3 &6000 &200 &0.005& 0.58& 0.058\\
32&B3 &6000 &200 &0.010& 0.58& 0.058\\
33&B3 &6000 &200 &0.050& 0.58& 0.058\\
34&B1 &6000 &500 &0.1 & 0.58& 0.058\\
35&B2 &6000 &500 &0.1& 0.58& 0.058\\
36&B3 &6000 &500 &0.1& 0.58& 0.058\\
& & &  & & & \\
\end{tabular}}
\end{table}
\vskip 0.5truecm
\noindent
col.1: simulation number\\
col.2: see Section 4.1.3\\
col.3: total  number of initial particles\\
col.4: total mass in unit of $10^{10}\,m\odot$\\
col.5: baryonic-to-total-mass ratio\\
col.6: initial triaxiality ratio (see Section 4.1.1)\\
col.7: initial spin parameter (see Section 4.1.1)\\

\clearpage
\centerline { Table 2. Properties at 15 Gyr for simulations with\\}
\centerline {      $M_{tot}=20$ and $M_{tot}/M_{bar}=0.1$ (see text) \phantom{000}}
\begin{table}[hbt]
\begin{tabular}{crrrrrrrr}
& & & & & & & & \\
IMF & $N_p$ & $r_{eff}$&
$r_D$ & $r_{opt}$ & $B\over D$ & ($M^*\over L_B$)($r_{opt}$) & ($M^*\over M_{DM}$)($r_{opt})$&f$_{gas}$\\
& & & & & & & &\\
B1 &6000 &3.23 &4.4& 12.8 &0.50& 3.2 & 0.58 & 0.34 \\
B2 &6000 &2.78 &3.5& 10.5 &0.70& 1.9 & 0.49& 0.62 \\
B3 &6000 &4.00 &4.7& 13.2&0.43& 7.1 & 0.55& 0.45 \\
B3 &2000 &3.03 &4.5& 13.1&0.47& 6.2 & 0.46& 0.41 \\
B3 &20000&4.18 &5.0& 12.8&0.44& 4.9 & 0.42& 0.58 \\
B3$^*$ &6000&2.50 &7.0&17.1&1.10& 16.1 & 0.50 & 0.34 \\
& & & & & &   & &\\
\end{tabular}
\end{table}
\vskip 0.5truecm
\noindent
$*$ for a model with $f_v$=0\\
col.1: see Section 4.1.3\\
col.2: total  number of initial particles\\
col.3: B effective radius (kpc)\\
col.4: exponential disk scale length (kpc) from  B surface brightness distribution\\
col.5: optical radius (kpc), i.e. radius where $L_{B}=0.83L_{B}^{tot}$ [54]\\
col.6: bulge-to-disk ratio from $x-y$ projection of B surface brightness distribution\\
col.7: stellar mass-to B-luminosity ratio inside optical radius\\
col.8: star-to-DM-mass ratio inside optical radius\\
col.9: residual gas fraction

\end{document}